\documentstyle[aps,twocolumn,prl,epsf]{revtex}

\begin{document}

\title{Inelastic neutron scattering peak in
Zn substituted YBa$_2$Cu$_3$O$_7$}

\author{ N.\ Bulut }

\address{
Department of Mathematics, Ko\c{c} University, 
Istinye, 80860 Istanbul, Turkey} 

\date{\today} 
\draft

\twocolumn[\hsize\textwidth\columnwidth\hsize\csname@twocolumnfalse\endcsname
\maketitle 

\begin{abstract}
The effects of nonmagnetic impurities on the neutron
scattering intensity are studied
for a model of the copper oxide layers
in the normal state.
The contribution to the 
${\bf Q}=(\pi,\pi)$ neutron scattering 
intensity from processes involving the scattering 
of the spin-fluctuations from an impurity 
with large momentum transfers 
are calculated
within the random phase approximation.
It is shown that 
this type of scatterings could lead to a peak in the 
neutron scattering intensity in the normal state.

\end{abstract}

]

An important feature found by the neutron scattering experiments 
on YBa$_2$Cu$_3$O$_7$ is a resonant peak which is observed
at 41 meV energy and ${\bf Q}=(\pi,\pi)$ 
momentum transfers \cite{Rossat-Mignod,Mook,Fong95}.
In this material, the peak is observed only 
in the superconducting state. 
The theories which have been put forward attribute 
the resonant peak
to an instability in the particle-particle \cite{Demler,Demler2}
or the magnetic channels \cite{Liu,Mazin,Bulut}.
It is of interest to determine how much each of these two channels 
contribute to the resonant peak in YBa$_2$Cu$_3$O$_7$.

Nonmagnetic Zn impurities have been used as 
a probe of the magnetic correlations in NMR experiments, 
and yielded valuable information \cite{Mahajan}. 
Neutron scattering experiments on Zn substituted 
YBa$_2$Cu$_3$O$_7$ also find that the Zn impurities have 
important effects on the 
spin fluctuation spectrum \cite{Sidis,Fong}.
Upon substituting 0.5$\%$ Zn, 
the resonant peak 
of the superconducting state becomes observable 
in the normal state as a broadened peak centered at 40 meV with a width
of 10 meV \cite{Fong}. 
As the temperature is lowered, 
the peak intensity increases smoothly without any 
observable discontinuity in the slope 
at the superconducting transition at $T_c=87^{\circ}$K.
It is found that significant amount of spectral weight 
develops in the peak already in the normal state. 
For instance, at $94^{\circ}$K the amplitude of the neutron
scattering intensity is about half of its value at $22^{\circ}$K.
In this sample, there is no observable spectral weight 
at frequencies less than $\sim 35$ meV.
However, when the concentration of the impurities 
is increased to $2\%$, 
the neutron scattering intensity develops weight at 
frequencies as low as 5 meV, and a broad
peak is observed at about 35 meV \cite{Sidis}. 
The momentum width of the structure at 35 meV 
is about 0.5 \AA$^{-1}$, which is twice that of the peak 
in the sample with 0.5$\%$ Zn. 
These are the main features of the changes 
induced by the Zn impurities.
Understanding the reason for why dilute Zn impurities induce a peak 
in the neutron scattering intensity in the normal state would 
be useful for understanding the origin of the resonant peak 
in pure YBa$_2$Cu$_3$O$_7$ \cite{Fong}.

Here, the effects of dilute Zn impurities on 
the neutron scattering spectral weight  
${\rm Im}\,\chi({\bf Q},\omega)$ will be studied for a 
two-dimensional model of the layered cuprates 
in the normal state.
The contributions to 
${\rm Im}\,\chi({\bf Q}=(\pi,\pi),\omega)$ from 
processes involving the scattering of the spin 
fluctuations from an impurity with large momentum 
transfers near $2{\bf k}_F$, where ${\bf k}_F$ is the 
Fermi momentum, are calculated within 
the random phase approximation (RPA).
It will be seen that this type of scatterings 
could lead to a peak in 
${\rm Im}\,\chi({\bf Q}=(\pi,\pi),\omega)$
in the normal state
at $\omega_0=2|\mu|$, where $\mu$ is the chemical potential.

During the scattering of the spin fluctuations 
from an impurity,
momentum is not conserved and,
hence, the ${\bf Q}=(\pi,\pi)$ component of 
the spin fluctuations could get mixed with 
the other wave vector components.
For instance,
if momentum ${\bf Q^*}$ is transfered during a 
scattering, then the 
${\bf Q}=(\pi,\pi)$ component of the spin fluctuations would get mixed 
with the ${\bf q}={\bf Q}-{\bf Q^*}$ component.
This process could be considered as an umklapp scattering 
of the spin fluctuations by the impurity potential.
It has been previously noted that this type of scatterings with large 
momentum transfers could play a role in enhancing the uniform 
susceptibility as antiferromagnetic correlations develop
in Zn substituted YBa$_2$Cu$_3$O$_{7-\delta}$ \cite{Bulut2}.
It has been also shown that without the umklapp scatterings, 
the main effect of the impuritites on 
${\rm Im}\,\chi({\bf Q},\omega)$ is to cause a smooth 
smearing \cite{Li}.
Here, it will be shown that, when included within RPA, 
the scattering of the spin fluctuations with momentum transfers
${\bf Q^*}\approx 2{\bf k}_F$
can lead to a peak in 
${\rm Im}\,\chi({\bf Q},\omega)$ near $\omega_0$
in the normal state.
The underlying reason for this is that there is 
a kinematic constraint against creating a particle-hole 
pair with center-of-mass momentum 
${\bf q}=(\pi,\pi)-2{\bf k}_F$ and energy 
$\omega > \omega_0$.
On the other hand, the scattering
of the spin fluctuations with momentum transfers  
away from $2{\bf k}_F$ has a smaller effect on 
${\rm Im}\,\chi({\bf Q},\omega\approx \omega_0)$.

In the following, 
the two-dimensional Hubbard model will be used to model 
the magnetic correlations of the pure system, and 
a static potential will be used to model 
the effective interaction
between the electrons and a nonmagnetic impurity.
The two-dimensional 
single-band Hubbard model is defined by 
\begin{eqnarray}
\label{Hubbard}
H=-t\sum_{\langle ij\rangle ,\sigma} 
(c^{\dagger}_{i\sigma}c_{j\sigma}
+c^{\dagger}_{j\sigma}c_{i\sigma})
+ U \sum_i 
c^{\dagger}_{i\uparrow} c_{i\uparrow} 
c^{\dagger}_{i\downarrow} c_{i\downarrow} \nonumber  \\
-\mu \sum_{i,\sigma}
c^{\dagger}_{i\sigma} c_{i\sigma}, 
\end{eqnarray}
where $c_{i\sigma}$ ($c^{\dagger}_{i\sigma}$)
annihilates (creates) an electron with spin $\sigma$
at site $i$,
$t$ is the near-neighbor hopping matrix element and 
$U$ is the onsite Coulomb repulsion.
For simplicity, the hopping $t$,
the lattice constant $a$ and $\hbar$ are set to 1.

When an impurity is introduced at site ${\bf r}_0$ 
of the system described by 
Eq.~({\ref{Hubbard}), the translational invariance is broken
and, in this case, the magnetic susceptibility is defined as
\begin{equation}
\chi({\bf q},{\bf q'},i\omega_m) = 
\int_0^{\beta} d\tau \, 
e^{i\omega_m \tau}
\langle 
m^{-}({\bf q},\tau)  m^{+}({\bf q'},0) 
\rangle,
\label{chi}
\end{equation}
where $m^{+}({\bf q})= N^{-1/2} \sum_{{\bf p}}
c^{\dagger}_{{\bf p}+{\bf q}\uparrow} 
c_{{\bf p}\downarrow}$,
$m^{-}({\bf q})=(m^+({\bf q}))^{\dagger}$, 
and $\omega_m=2m\pi T$. 
The susceptibility defined by 
Eq.~(\ref{chi}) 
has both diagonal ${\bf q}={\bf q'}$ and 
off-diagonal ${\bf q}\neq{\bf q'}$ terms.
If a static effective interaction 
is assumed between the impurity at
site ${\bf r}_0$ and the electrons,
then the RPA expression for 
$\chi$ in the Hubbard model becomes 
\begin{eqnarray} 
\label{rpa}
\chi({\bf q},{\bf q}',i\omega_m) = & &
\chi_0({\bf q},{\bf q}',i\omega_m) \nonumber \\
+ & &
U \sum_{{\bf q}''}
\chi_0({\bf q},{\bf q}'',i\omega_m)
\chi({\bf q}'',{\bf q}',i\omega_m),
\end{eqnarray}
where
$\chi_0({\bf q},{\bf q'},i\omega_m)$ is the irreducible 
susceptibility dressed with the 
impurity scatterings, but not with the Coulomb correlations. 
Here,
the diagonal ${\bf q}={\bf q'}$
terms of the irreducible susceptibility 
$\chi_0({\bf q},{\bf q'},i\omega_m)$ 
will be approximated 
by the Lindhard function of the pure system,
$\chi^L_0({\bf q},i\omega_m)$.
This is a good approximation 
since it is already known from 
previous calculations that the effect of impurity scattering 
on the diagonal irreducible susceptibility is to cause a
smooth suppression \cite{Li},
and this has a small effect compared to the 
nearly singular contribution arising from the 
umklapp processes.
The 
${\bf q}\neq{\bf q'}$ off-diagonal terms of 
$\chi_0({\bf q},{\bf q'},i\omega_m)$
will be calculated in the lowest order in the strength of the 
effective electron-impurity interaction as shown
diagrammatically in Fig.~1(a).
It will be seen that, within this model,
the impurity induced changes in the 
${\bf Q}=(\pi,\pi)$ spin-fluctuation spectrum 
are closely related to the 
spectrum of the 
irreducible off-diagonal susceptibility 
$\chi_0({\bf Q},{\bf q},\omega)$ with 
${\bf q}={\bf Q}-2{\bf k}_F$.
After including the effects of the impurity scatterings 
and the Coulomb correlations,
the impurity averaging 
is done by averaging over the impurity location ${\bf r}_0$, 
which sets ${\bf q}={\bf q'}$.
This way, one obtains 
$\chi({\bf Q},{\bf Q},i\omega_m)$,
which gives the neutron scattering spectral weight 
after analytic continuation
$i\omega_m\rightarrow \omega+i\delta$.
In the following, 
$\chi({\bf Q},{\bf Q},i\omega_m)$ will be denoted by 
$\chi({\bf Q},i\omega_m)$.

A static effective electron-impurity interaction for an 
impurity located at site ${\bf r}_0$ can be written as 
\begin{equation}
V_{eff} =
\label{V}
{1\over N}\sum_{{\bf k}}
e^{i{\bf k}\cdot {\bf r}_0}
V_{{\bf k}} \sum_{{\bf p},\sigma} 
c^{\dagger}_{{\bf p}+{\bf k}\sigma} c_{{\bf p}\sigma},
\end{equation}
which involves scatterings 
with all possible wave vectors.
It has been noted that the electronic correlations play a role 
in determining the structure of $V_{\bf k}$ \cite{Ziegler}.
It will be seen that 
for $\omega\approx \omega_0$
the important scatterings are the 
\begin{figure} 
\begin{center}
\leavevmode
\epsfxsize=6cm 
\epsfysize=2.5cm 
\epsffile{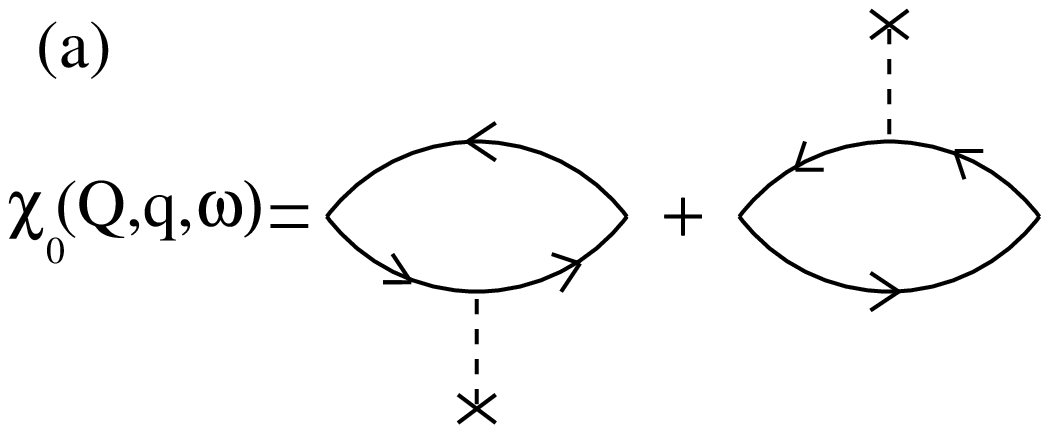}
\end{center}
\begin{center}
\leavevmode
\epsfxsize=6cm 
\epsfysize=4cm 
\epsffile[18 205 592 598]{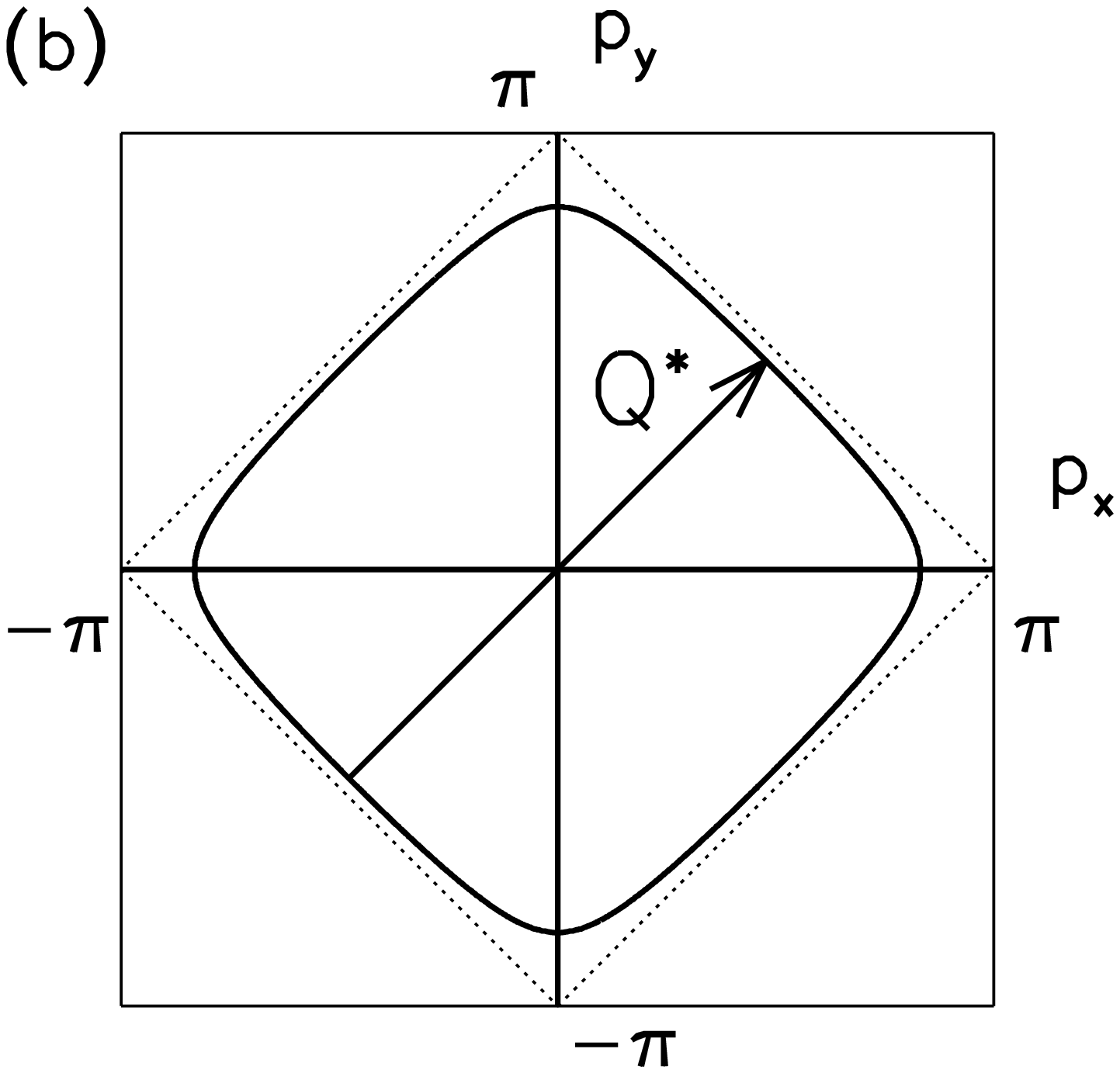}
\end{center}
\caption{
(a) Feynman diagrams representing the lowest order terms 
of the irreducible umklapp susceptibility 
$\chi_0({\bf Q},{\bf q},\omega)$.
(b) Sketch of a single-particle scattering process 
involving ${\bf Q^*}= 2{\bf k}_F$ momentum transfer
in the Brillouin zone.
Here
${\bf k}_F$ is the Fermi wave vector along $(1,1)$
and 
the solid curve represents the Fermi surface of 
the noninteracting system at finite hole doping.
}
\label{fig1}
\end{figure}
\noindent 
ones which 
transfer ${\bf Q^*}=2{\bf k}_F$, since in this case
$\chi_0({\bf Q},{\bf q},\omega)$ 
with ${\bf q}={\bf Q}-{\bf Q^*}$
has nearly singular behaviour.
For other values of ${\bf Q^*}$, 
$\chi_0({\bf Q},{\bf q},\omega)$ 
is a smooth function of 
$\omega$ with small amplitude.
Hence, Eq.~(\ref{rpa}) will be solved by using only the 
${\bf Q^*}$ component of Eq.~(\ref{V}),
\begin{equation}
\label{V2}
V_0\sum_{{\bf p}\sigma} 
c^{\dagger}_{{\bf p}+{\bf Q^*}\sigma}
c_{{\bf p}\sigma},
\end{equation}
where $V_0=V_{\bf Q^*}$ is taken as a parameter.
Since the important umklapp contributions originate 
from a small range of values of ${\bf Q^*}$ near $2{\bf k}_F$, this
is a reasonable approximation. 
This procedure 
for solving Eq.~(\ref{rpa})
was necessary, because the calculation needs
to be carried out on a large lattice in order 
to have sufficient frequency resolution.
A single-particle scattering process 
given by
Eq.~(\ref{V2}) is illustrated in Fig.~1(b)
for ${\bf Q^*}=2{\bf k}_F$ momentum transfer
in the Brillouin zone.

Within the presence of the scattering term given by 
Eq.~(\ref{V2}), the solution of the RPA expression,
Eq.~(\ref{rpa}), becomes
\begin{eqnarray} 
\label{chi22}
\chi({\bf Q},\omega) = 
D^{-1}
\bigg\{
\chi^L_0({\bf Q},&&\omega) 
(1 - U\chi^L_0({\bf q},\omega)) \nonumber \\
&& + 
4 U(\chi_0({\bf Q},{\bf q},\omega))^2 
\bigg\},
\end{eqnarray}
where 
$D=(1-U\chi^L_0({\bf Q},\omega))
(1 - U\chi^L_0({\bf q},\omega)) 
- 4 (U\chi_0({\bf Q},{\bf q},\omega))^2$
and ${\bf q}={\bf Q}-{\bf Q^*}$.
Here,
the factor of 4 multiplying 
$(\chi_0({\bf Q},{\bf q},\omega))^2$ 
is to take into account the 
scatterings with momentum transfers 
$(\pm Q^*,\pm Q^*)$ and 
$(\pm Q^*,\mp Q^*)$, where ${\bf Q^*}=(Q^*,Q^*)$. 
The expression for 
the irreducible umklapp susceptibility 
$\chi_0({\bf Q},{\bf q},i\omega_m)$, which is 
illustrated in Fig.~1(a), is given by
\begin{eqnarray} 
\label{chi012}
\chi_0({\bf Q},{\bf q},i\omega_m) = 
-V_0 {T\over N} \sum_{{\bf p},i\omega_n} 
\bigg\{ 
G_0&&({\bf p}+{\bf Q},i\omega_n+i\omega_m) \nonumber \\
\times
G_0({\bf p},i\omega_n) 
&&G_0({\bf p}+{\bf Q}-{\bf q},i\omega_n) \nonumber \\
+ 
G_0({\bf p},i\omega_n) G_0({\bf p}+{\bf q},&&i\omega_n+i\omega_m) 
\nonumber \\
\times
G_0({\bf p}&&+{\bf Q},i\omega_n+i\omega_m)
\bigg\}.
\end{eqnarray}
Here,
$G_0({\bf p},i\omega_n)=(i\omega_n-\varepsilon_{\bf p})^{-1}$
with
$\omega_n=(2n+1)\pi T$ and 
$\varepsilon_{\bf p} = -2t (\cos{p_x} + \cos{p_y}) - \mu$
is the single-particle Green's function
of the pure system.
After summing over $i\omega_n$ in Eq.~(\ref{chi012}) analytically,
the resulting sum over ${\bf p}$ is carried out numerically.
The Lindhard susceptibility $\chi_0^L$ 
used in Eq.~(\ref{chi22}) is given by 
\begin{equation}
\label{Lindhard}
\chi^L_0({\bf k},\omega) = 
{1\over N}
\sum_{\bf p}
{
f(\varepsilon_{{\bf p}+{\bf k}}) -
f(\varepsilon_{\bf p}) \over 
\omega - ( \varepsilon_{{\bf p}+{\bf k}} 
- \varepsilon_{\bf p} ) + i\delta},
\end{equation}
which is also evaluated numerically.
The replacement of the diagonal irreducible susceptibility 
by the Lindhard susceptibility in solving Eq.~(\ref{rpa})
does not affect the results presented here.
It has been shown in 
Ref.~\cite{Li} that scatterings from an onsite impurity potential 
lead to a smooth smearing of the structure in 
the diagonal irreducible susceptibility.
A similar effect was found for 
the $\omega=0$ case 
in Ref.~\cite{Bulut2}
using an extended impurity potential.

The following results have been obtained by using 
$U=1.7$, temperature $T=0.05$ 
and filling $\langle n\rangle=0.86$.
For these values of $T$ and $\langle n\rangle$,
one has $\omega_0=2|\mu| \approx 0.55$.
Figure~2 shows results obtained for momentum transfer
${\bf Q^*}=2{\bf k}_F$, where ${\bf k}_F$ was taken 
along (1,1) for simplicity. 
In Fig.~2(a), the real and the imaginary parts 
of $\chi_0({\bf Q},{\bf q},\omega)$ are shown. 
For comparison, in Fig.~2(b), 
the $\omega$ dependence of the Lindhard susceptibility 
$\chi^L_0({\bf Q},\omega)$ is shown. 
The solid line in Fig.~2(c)
represents
${\rm Im}\,\chi({\bf Q},\omega)$
for $V_0=0.05$.
Also shown in this figure by the dashed line is the result 
for $V_0=0$,
which corresponds to the pure case.
Here, it is seen that 
${\rm Im}\,\chi_0({\bf Q},{\bf q},\omega)$ is finite 
for $\omega< \omega_0$ with a sharp cut-off at $\omega_0$.
The cut-off in ${\rm Im}\,\chi_0({\bf Q},{\bf q},\omega)$ 
is due to the fact that there is a kinematic 
constraint against creating a particle-hole pair with the 
center-of-mass momentum ${\bf q}=(\pi,\pi)-2{\bf k}_F$ and energy 
$\omega>\omega_0$.
In the corresponding 
real part there 
is a sharp peak at $\omega_0$.
Both the real and the imaginary parts of 
$\chi_0({\bf Q},{\bf q},\omega)$ play a role in suppressing 
the denominator of the expression in Eq.~(\ref{chi22}), 
and in leading to the peak in 
${\rm Im}\,\chi({\bf Q},\omega)$
at $\omega_0$.
In this model, the peak is originating from the magnetic channel,
since it is due to an instability of Eq.~(\ref{rpa}).

The spectrum of $\chi_0({\bf Q},{\bf q},\omega)$ seen in 
Fig.~2(a) is quite different than that of 
the Lindhard susceptibility
$\chi_0({\bf Q},\omega)$ seen in Fig.~2(b).
In the limit $T\rightarrow 0$, 
${\rm Im}\,\chi_0({\bf Q},\omega)$  
develops a gap for $\omega<\omega_0$.
On the other 
\begin{figure} 
\begin{center}
\leavevmode
\epsfxsize=8cm 
\epsfysize=9cm 
\epsffile{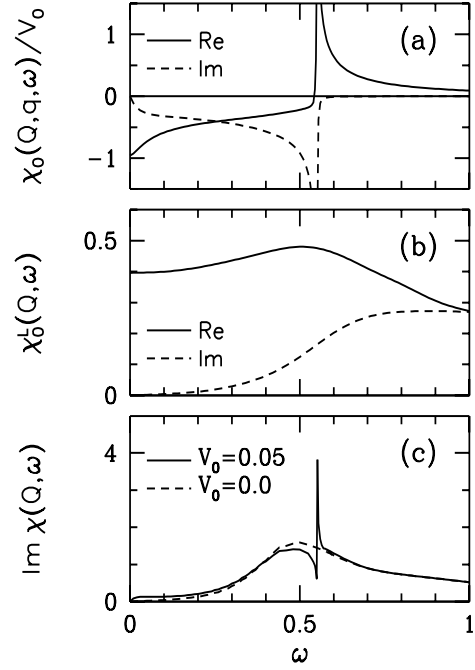}
\end{center}
\caption{
(a) Irreducible umklapp susceptibility 
$\chi_0({\bf Q},{\bf q},\omega)$
versus $\omega$
normalized by $V_0$. 
Here ${\bf q}={\bf Q}-{\bf Q^*}$.
(b) Lindhard susceptibility 
$\chi^L_0({\bf Q},\omega)$ versus $\omega$.
(c) Spin-fluctuation spectral weight 
${\rm Im}\,\chi({\bf Q},\omega)$ 
versus $\omega$ for $V_0=0.05$ and 0.0.
These results were obtained for 
${\bf Q^*}=2{\bf k}_F$.
}
\label{fig2}
\end{figure}
\noindent
hand, 
${\rm Im}\,\chi_0({\bf Q},{\bf q},\omega)$ is finite only for 
$\omega<\omega_0$.
Hence, 
the umklapp scatterings could contribute to 
${\rm Im}\,\chi({\bf Q},\omega)$ for 
$\omega<\omega_0$.
The quantitative features of 
${\rm Im}\,\chi({\bf Q},\omega)$ 
depend on the model parameters.
For instance, if $U$ is increased from 1.7 to 2.0,
the hump in ${\rm Im}\,\chi({\bf Q},\omega)$ 
centered at $\omega=0.5$ grows and the dip below the peak
gets smaller.
The scattering of the quasiparticles 
by the spin fluctuations would also 
affect the quantitative features.

Figures 3(a) and (b) show the temperature evolution of 
$\chi_0({\bf Q},{\bf q},\omega)$ and
${\rm Im}\,\chi({\bf Q},\omega)$
for $V_0=0.05$ on a smaller $\omega$ scale.
As $T$ increases, the peak in 
${\rm Im}\,\chi({\bf Q},\omega)$ disappears.
Figure 3(c) shows the development of the peak in 
${\rm Im}\,\chi({\bf Q},\omega)$ as $V_0$ is turned on
at $T=0.05$.

In order to have further information on the momentum and 
frequency structure of 
$\chi_0({\bf Q},{\bf q},\omega)$, 
results on its real part are shown in Fig.~4.
Figure 4(a) shows 
${\rm Re}\,\chi_0({\bf Q},{\bf q},\omega)$ 
versus $\omega$ obtained at 
values of ${\bf Q^*}$ which are slightly different 
than $2{\bf k}_F$,
but still along (1,1).
Here, the numbers next to the curves indicate 
the magnitude of ${\bf Q^*}$ in units of 
$2k_F$.
As ${\bf Q^*}$ is varied away from 
$2{\bf k}_F$, the resonance frequency shifts and the amplitude of 
${\rm Re}\,\chi_0({\bf Q},{\bf q},\omega)$ decreases.
Hence, when all the contributions from 
scatterings with momentum transfers 
near $2{\bf k}_F$ are taken into account, 
one expects the peak in 
${\rm Im}\,\chi({\bf Q},\omega)$ at $\omega_0$ to have 
a finite width.
However,
it is not possible to estimate the total spectral weight in 
the peak.
For instance, the detailed ${\bf k}$ dependence of 
$V_{\bf k}$ for ${\bf k}\sim 2{\bf k}_F$ is 
not known.
\begin{figure} 
\begin{center}
\leavevmode
\epsfxsize=8cm 
\epsfysize=9cm 
\epsffile{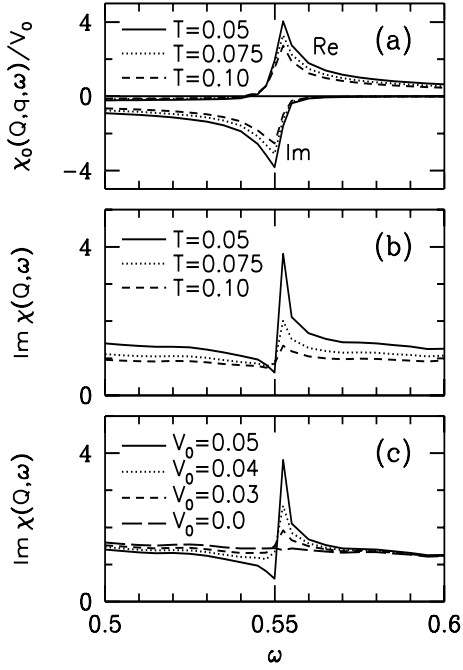}
\end{center}
\caption{
Temperature variation of 
(a) $\chi_0({\bf Q},{\bf q},\omega)/V_0$ 
and 
(b) ${\rm Im}\,\chi({\bf Q},\omega)$ versus $\omega$
for $V_0=0.05$.
In these two figures,
the chemical potential has been kept fixed, while increasing
the temperature.
(c) ${\rm Im}\,\chi({\bf Q},\omega)$
versus $\omega$ for various values of $V_0$
at $T=0.05$.
These results were obtained for 
${\bf Q^*}=2{\bf k}_F$.
}
\label{fig3}
\end{figure}
\begin{figure} 
\begin{center}
\leavevmode
\epsfxsize=8.5cm 
\epsfysize=10cm 
\epsffile{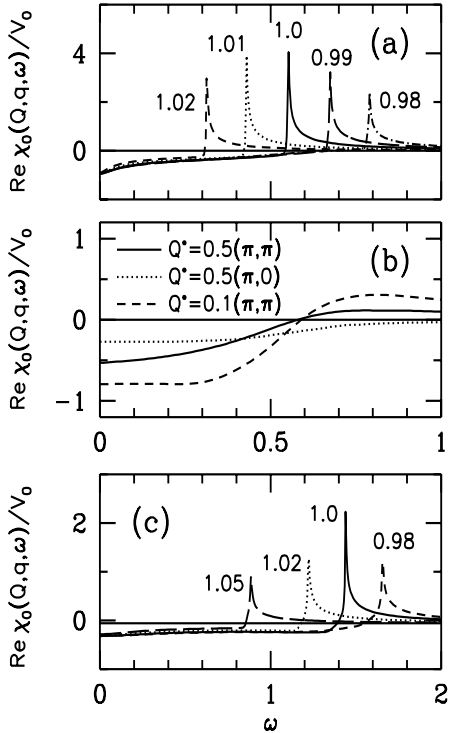}
\end{center}
\caption{
${\rm Re}\,\chi_0({\bf Q},{\bf q},\omega)/V_0$ versus $\omega$
(a) for ${\bf Q^*}$ near $2{\bf k}_F$ and 
(b) for various values of ${\bf Q^*}$ away from $2{\bf k}_F$ 
at $\langle n\rangle =0.86$.
(c) ${\rm Re}\,\chi_0({\bf Q},{\bf q},\omega)/V_0$ versus $\omega$
for ${\bf Q^*}$ near $2{\bf k}_F$ at
$\langle n\rangle=0.7$.
}
\label{fig4}
\end{figure}

In Fig. 4(b), 
${\rm Re}\,\chi_0({\bf Q},{\bf q},\omega)$
is shown for various values of ${\bf Q^*}$ away from 
$2{\bf k}_F$.
Here one observes that 
${\rm Re}\,\chi_0({\bf Q},{\bf q},\omega)$
is a smooth function of $\omega$ and has a much
smaller amplitude near $\omega_0$ than 
in the case for ${\bf Q^*}\approx 2{\bf k}_F$.
Other values for ${\bf Q^*}$ away from $2{\bf k}_F$ were also 
tried, but results similar to those in 
Fig.~4(b) were obtained.
This supports keeping 
only the ${\bf Q^*}\approx 2{\bf k}_F$ component, Eq.~(\ref{V2}),
of the effective impurity interaction, Eq.~(\ref{V}),
in solving the RPA equation for $\chi({\bf Q},\omega)$ 
at $\omega$ near $\omega_0$.
Finally, Fig.~4(c) shows 
${\rm Re}\,\chi_0({\bf Q},{\bf q},\omega)$
versus $\omega$ for ${\bf Q^*}$ near $2{\bf k}_F$
at filling $\langle n\rangle=0.7$, where
the corresponding $\omega_0$ is 1.44.
Comparing with the results shown in Fig.~4(a), 
one observes that the qualitative features 
do not change with the band filling.

The neutron scattering experiments 
by Fong {\it et al.} \cite{Fong}
have found that, as the temperature is lowered 
below $250^{\circ}$K, a peak develops 
at $\omega=40$ meV
in ${\rm Im}\,\chi({\bf Q},\omega)$ 
of 0.5\% Zn substituted YBa$_2$Cu$_3$O$_7$.
Here, the effects of dilute nonmagnetic impurities 
on ${\rm Im}\,\chi({\bf Q},\omega)$ 
have been studied within a two-dimensional model
in the normal state.
It has been found that the impurity induced changes in 
${\rm Im}\,\chi({\bf Q},\omega)$ are closely related 
to the spectrum of the 
irreducible off-diagonal susceptibility
$\chi_0({\bf Q},{\bf q},\omega)$.
The possibility of the structure in 
$\chi_0({\bf Q},{\bf q},\omega)$ 
at $\omega\approx\omega_0$
leading to a peak in ${\rm Im}\,\chi({\bf Q},\omega)$ has been pointed out. 
Within this model, the peak is due to the 
impurity scattering of the spin fluctuations with momentum
transfers near $2{\bf k}_F$.
However, in obtaining these results 
the Coulomb correlations 
were treated within the random phase approximation, and 
only the lowest order terms 
of $\chi_0({\bf Q},{\bf q},\omega)$ have been included.
It remains to be seen by how much the higher order corrections will 
modify the results discussed here.

The author thanks P. Bourges, H.F. Fong and B. Keimer
for helpful discussions.
The numerical computations reported in this paper were performed 
at the Center for Information Technology at Ko\c{c} University.


    

\begin{thebibliography}{999}

\bibitem{Rossat-Mignod} J. Rossat-Mignod {\it et al.}, 
Physica {\bf 185-189C}, 86 (1991).

\bibitem{Mook} H.A. Mook {\it et al.},
\prl {\bf 70}, 3490 (1993).

\bibitem{Fong95} H.F. Fong {\it et al.},
\prl {\bf 75} 316 (1995).

\bibitem{Demler} E. Demler {\it et al.}, 
\prl {\bf 75}, 4126 (1995).

\bibitem{Demler2} E. Demler {\it et al.}, 
\prb {\bf 58}, 5719 (1998).

\bibitem{Liu} D.Z. Liu {\it et al.}, 
\prl {\bf 75}, 4130 (1995).

\bibitem{Mazin} I.I. Mazin {\it et al.}, 
\prl {\bf 75}, 4134 (1995).

\bibitem{Bulut} N. Bulut {\it et al.}, 
\prb {\bf 53}, 5149 (1996).

\bibitem{Mahajan} A.V. Mahajan {\it et al.},
\prl {\bf 72}, 3100 (1994).

\bibitem{Sidis} Y. Sidis {\it et al.}, 
\prb {\bf 53}, 6811 (1996).

\bibitem{Fong} H.F. Fong {\it et al.},
\prl {\bf 82}, 1939 (1999).

\bibitem{Bulut2} N. Bulut, preprint, 
cond-mat/9906185.

\bibitem{Li} J.-X. Li {\it et al.},
\prb {\bf 58}, 2895 (1998).

\bibitem{Ziegler} W. Ziegler {\it et al.}, 
\prb {\bf 53}, 8704 (1996).

\end{thebibliography}
\end{document}